\documentclass[twocolumn]{aastex631}
\usepackage[T1]{fontenc}
\usepackage{amsmath,amssymb}
\usepackage{graphicx}
\usepackage{natbib}
\usepackage{microtype}
\usepackage{placeins}

\newcommand{\Mdot}{\dot{M}}
\newcommand{\MEdd}{\dot{M}_{\rm Edd}}
\newcommand{\Qvis}{Q^{+}}
\newcommand{\Qrad}{Q^{-}_{\rm rad}}
\newcommand{\Qadv}{Q^{-}_{\rm adv}}
\newcommand{\Qwin}{Q^{-}_{\rm w}}
\newcommand{\btor}{\beta_{\varphi}}
\newcommand{\bpol}{\beta_{z}}
\newcommand{\bgas}{\beta_{\rm g}}
\newcommand{\brad}{\beta_{\rm rad}}
\newcommand{\ptot}{p_{\rm tot}}
\newcommand{\prad}{p_{\rm rad}}
\newcommand{\pgas}{p_{\rm gas}}
\newcommand{\OmegaK}{\Omega_{\rm K}}
\newcommand{\fadv}{f_{\rm adv}}
\newcommand{\fwin}{f_{\rm w}}
\newcommand{\rg}{r_{\rm g}}
\newcommand{\gp}{\gamma_{p}}
\newcommand{\gammaeq}{\gamma_{\rm eq}}
\newcommand{\etaz}{\eta_{\zeta}}
\newcommand{\Gnf}{\Gamma_{\rm nf}}
\newcommand{\Dchi}{D_{\chi}}
\newcommand{\Cchi}{C_{\chi}}
\newcommand{\etad}{\eta_{\delta}}
\newcommand{\Sd}{\mathcal{S}_{\delta}}
\newcommand{\Sc}{\mathcal{S}_{\rm crit}}
\newcommand{\lmdot}{\lambda_{\dot M}}

\shorttitle{Thermal Stability with Net Vertical Flux}
\shortauthors{Abbassi et al.}
\defcitealias{Habibi2019}{H19}

\begin{document}

\title{Thermal Stability of Radiation-Pressure-Dominated Accretion Disks Threaded by Net Vertical Magnetic Flux}

\author[0000-0003-0428-2140]{Shahram Abbassi}
\affiliation{Department of Physics and Astronomy, Western University,
London, Ontario N6A 3K7, Canada}
\email{sabbassi@uwo.ca}
\correspondingauthor{Shahram Abbassi}

\author{Armin Memarian}
\affiliation{Department of Physics, Faculty of Sciences,
Ferdowsi University of Mashhad,
Mashhad 91775-1436, Iran}
\author[0000-0002-5730-0376]{Samik Mitra}
\affiliation{International Centre for Theoretical Sciences, Tata Institute of Fundamental Research,
Bengaluru 560089, India}
\email{samik.mitra@icts.res.in}

\begin{abstract}
The classical radiation-pressure instability predicts strong thermal variability in luminous black-hole accretion disks, whereas most disk-dominated X-ray binary soft states remain comparatively stable. We examine whether net vertical magnetic flux can weaken this instability through its contribution to the radial stress. The stability depends not only on the equilibrium magnetic stress but also on how that stress changes during a thermal perturbation. We write the local stability condition in terms of the logarithmic heating response, $q_+<q_{+,\rm crit}$, which avoids specifying how the turbulent stress is divided into reference and net-flux components. For the illustrative closure $\delta=\zeta\sqrt{\bpol\btor}/\alpha$, $\etad=-d\ln\delta/d\ln H$ describes the response of the fractional stress correction, while $\Gnf=d\ln W_{\rm nf}/d\ln H$ describes the response of the additional net-flux stress itself. For an accretion disk around a $10M_\odot$ black hole at $R=20\rg$ and $\Mdot=0.5\MEdd$, we adopt fixed local mass flux on the thermal timescale, $(\gamma,\gp)=(1,0)$, and $d\ln\zeta/d\ln H=0$. Marginal stability then occurs at $\bpol^{\rm crit}=0.0176$ for $\zeta=1$ and $0.1338$ for $\zeta=0.25$. These thresholds depend on the adopted stress closure and field-response prescription and should not be interpreted as universal magnetic-pressure fractions. Fixed-$B_z$ equilibrium sequences, using a separate relation for the variation of $B_\varphi$ between steady states, show that increasing vertical field narrows the thermally unstable accretion-rate interval but does not eliminate it for either $H/R<0.1$ or $H/R<0.2$. The relevant quantity for stabilization is therefore the thermal response of the stress associated with net vertical flux rather than the vertical magnetic-pressure fraction alone.
\end{abstract}


\section{Introduction}
\label{sec:intro}

Geometrically thin, optically thick accretion disks provide the
standard framework for radiatively efficient black-hole accretion
\citep{Shakura1973,Novikov1973}. They account for the dominant thermal
component of black-hole X-ray binaries (BHXBs) in the high/soft state
\citep{Gierlinski2004,Done2007} and for much of the optical and ultraviolet
emission from luminous active galactic nuclei (AGN) \citep{Malkan1982}.
Despite this success, the classical model faces a longstanding difficulty in
its radiation-pressure-dominated inner region. If the turbulent stress scales
with the total pressure, a positive temperature perturbation increases viscous
heating more rapidly than radiative cooling, producing thermal instability on
approximately the local thermal timescale
\citep{Lightman1974,Shakura1976,Piran1978}.

This prediction is difficult to reconcile with observations. Most
disk-dominated BHXB soft states remain remarkably steady over a broad
luminosity range \citep{Gierlinski2004,Done2007}, whereas strong recurrent
variability resembling a radiation-pressure limit cycle is confined to a small
number of systems, most notably GRS~1915$+$105
\citep{Done2007,McClintock2006}. Explaining both the stability of most luminous
thin disks and the persistence of large-amplitude variability in a minority of
sources remains an important problem in accretion theory.

Several mechanisms have been proposed to weaken the thermal mode,
including stresses weighted toward the gas pressure \citep{Sakimoto1981},
coronal dissipation
\citep{Svensson1994,Kawanaka2020,Tajmohammadi2025}, radial advection
\citep{Abramowicz1988}, magnetically elevated or supported disks
\citep{BegelmanPringle2007,Zhu2013}, and the loss of energy or angular
momentum through disk winds \citep{Li2014,Wu2022,Cao2022}. More generally,
the stability boundary changes if the turbulent stress responds dynamically
to a thermal perturbation rather than maintaining a fixed proportionality to
the local pressure. \citet{Naddaf2026}, for example, introduced a
thermodynamic dependence of the effective viscosity parameter. Here we
consider a different possibility in which the modified heating response is
associated with the stress generated in the presence of net vertical magnetic
flux.

Radiation-MHD simulations also show that the thermal outcome depends
on the magnetic and radiative structure of the flow. \citet{Hirose2009} found
no runaway over approximately forty cooling times and reported that stress
fluctuations preceded pressure fluctuations, whereas \citet{Jiang2013}
obtained eventual expansion or collapse in sufficiently large shearing boxes.
These results indicate that domain size, magnetic topology, radiative
transport, vertical structure, and stress--pressure timing all influence the
local thermal mode. They also suggest that equilibrium magnetic pressure
alone is not sufficient to characterize stability.

Net vertical magnetic flux is particularly relevant because it changes
the saturation of the magnetorotational instability and modifies both radial
and vertical magnetic stresses
\citep{Hawley1995,BaiStone2013,Salvesen2016,Sadowski2016}. Global
radiation-MHD calculations have produced magnetically supported inner disks
with substantial vertical structure \citep{Jiang2019Sub}, as well as luminous
configurations that remain stable or only weakly variable under strong
magnetization \citep{Mishra2022}. Recent radiative GRMHD simulations further
show that magnetic support, advection, and complex vertical structure can
coexist over a broad range of accretion rates
\citep{Zhang2025,Lancova2026}. The diversity of these solutions argues
against a universal stability threshold expressed solely in terms of
$B_z^2/(8\pi\ptot)$. A more direct diagnostic is the response of the heating
rate to a thermal perturbation.

Analytical models containing a toroidal magnetic field provide an
important starting point. \citet{Zheng2011} showed that an MRI-generated
$B_\varphi$ can weaken the radiation-pressure instability, and
\citet{Habibi2019} extended the analysis to include a parametric disk wind.
The remaining issue is how to incorporate ordered vertical flux without
identifying its equilibrium pressure contribution with its effect on the
perturbed radial stress. The radial Maxwell stress is proportional to
$-B_RB_\varphi$, not directly to $B_zB_\varphi$, and the relation between the
imposed vertical field, the coherent radial field, and the resulting stress
depends on the magnetic geometry and nonlinear MRI saturation. The
equilibrium stress enhancement and its thermal response therefore need not be
controlled by the same parameter.

We treat these two effects separately. We denote the fractional
net-flux contribution to the stress by $\delta$ and define
\begin{equation}
\Sd\equiv \frac{\delta}{1+\delta}\etad,
\qquad
\etad\equiv-\left.\frac{d\ln\delta}{d\ln H}\right|_{\Sigma,R}.
\label{eq:Sintro}
\end{equation}
For a pressure-scaled reference stress, $\Sd$ measures how strongly the net-flux contribution reduces the temperature dependence of the dissipation. The same stability condition can also be written directly in terms of the total logarithmic heating slope, without separating the turbulent stress into reference and net-flux components. This form is more general and can be compared with radiation-MHD simulations when the heating response is measured over the relevant thermal timescale. In turbulent flows, phase lags, frequency dependence, and nonlocal transport can affect this comparison and should be taken into account. In Eq. \ref{eq:Sintro}, $\etad$ is the response exponent of the fractional stress correction $\delta$; it is not the response exponent of the additional net-flux stress $W_{\rm nf}$.

In this work, we derive the local thermal-stability condition for a thin disk containing an MRI-generated toroidal field, ordered vertical flux, advective cooling, and an optional parametric wind. The vertical and toroidal fields are allowed to respond differently during the perturbation, while the equilibrium enhancement of the radial stress is kept distinct from its thermal derivative. We then adopt a phenomenological magnetic-stress prescription to express the criterion as a critical vertical magnetic-pressure fraction. To examine equilibrium families, we construct wind-free fixed-$B_z$ sequences and introduce a separate exponent for the variation of the saturated toroidal field between distinct steady states. The local thermal response and the equilibrium $B_\varphi(H)$ relation are therefore not identified with one another. This comparison distinguishes the stability of an individual equilibrium state from the persistence of an unstable branch along an entire sequence. Possible applications to BHXB variability and changing-look AGN remain qualitative. The model does not calculate magnetic-flux transport, transition times, nonlinear amplitudes, recurrence times, or duty cycles, so these systems are used only as physical motivation rather than as quantitative predictions.

Section~\ref{sec:model} introduces the disk structure, magnetic-field
evolution, and stress prescription. Section~\ref{sec:stability} derives the
local thermal-stability condition. Section~\ref{sec:results} presents the
critical stress response, the corresponding vertical magnetic fractions, and the fixed-$B_z$ equilibrium sequences. Section~\ref{sec:discussion} discusses
the physical interpretation, comparison with radiation-MHD simulations, and
the limitations of the local model.

\section{Model}
\label{sec:model}

We consider a steady, axisymmetric, optically thick disk in a Newtonian
potential. The angular velocity is Keplerian,
$\OmegaK=(GM/R^3)^{1/2}$, the gravitational radius is
$\rg=GM/c^2$, and the inner boundary is $R_{\rm in}=6\rg$. The fiducial
calculation is performed at $R=20\rg$. We define
\begin{equation}
\MEdd\equiv \frac{L_{\rm Edd}}{0.1c^2}.
\label{eq:medd}
\end{equation}

The total midplane pressure is
\begin{equation}
\ptot=\prad+\pgas+p_\varphi+p_z,
\label{eq:eos}
\end{equation}
where
$\prad=aT^4/3$,
$\pgas=\rho k_{\rm B}T/(\mu m_{\rm H})$,
$\rho=\Sigma/(2H)$,
$\mu=0.615$,
$p_\varphi=B_\varphi^2/(8\pi)$, and
$p_z=B_z^2/(8\pi)$. We write $\beta_i=p_i/\ptot$, so that
\begin{equation}
\brad+\bgas+\btor+\bpol=1,
\end{equation}
and define the total magnetic-pressure fraction as
$\beta_{\rm mag}\equiv\btor+\bpol$.

A vertically uniform $B_z$ does not contribute a vertical
magnetic-pressure gradient. The pressure entering the hydrostatic balance is
therefore written as
\begin{equation}
p_{\rm s}=\ptot-(1-\chi)p_z
=\frac{\Sigma\OmegaK^2H}{2}.
\label{eq:support}
\end{equation}
The main calculation adopts the strict thin-disk convention $\chi=0$.
The alternative choice $\chi=1$ includes the vertical magnetic pressure in
the scalar support pressure and is used only as a sensitivity test.

The magnetic-field responses during a local thermal perturbation are
parameterized by
\begin{align}
d\ln B_\varphi&=-\gamma\,d\ln H,
\label{eq:btor}\\
d\ln B_z&=-\gp\,d\ln H.
\label{eq:bpol}
\end{align}

The exponents $\gamma$ and $\gp$ are effective response indices on the
thermal timescale, rather than purely geometrical dilution parameters. The
choice $\gp=0$ represents frozen vertical flux during the local perturbation.
The fiducial value $\gamma=1$ corresponds to $B_\varphi\propto H^{-1}$ only
when geometric expansion dominates over dynamo regeneration, magnetic
buoyancy, and reconnection. Radiation-MHD simulations show that stress and
pressure need not respond simultaneously, while net-flux MRI simulations show
that the coherence and cyclic behavior of the toroidal field change with
magnetization \citep{Hirose2009,Salvesen2016}. The explored range of $\gamma$
therefore brackets uncertain dynamical responses of the toroidal field and
should not be interpreted as a simulation-calibrated geometrical prior.

The calculation does not determine $\gp$ from the mass-accretion time because magnetic flux may advect or diffuse at a speed different from that of the gas \citep{Lubow1994,Beckwith2009}. The exponent $\gamma$ is reserved for the local thermal derivative. For the equilibrium families in Section~\ref{sec:results}, we introduce $\gammaeq$ through $B_\varphi H^{\gammaeq}={\rm const}$ and adopt $\gammaeq=1$ only for the illustrative fixed-$B_z$ sequences. The numerical choice $\gamma=\gammaeq=1$ does not identify the two derivatives.

We take $W>0$ to denote the magnitude of the vertically integrated stress that transports angular momentum outward. The phenomenological net-flux correction is therefore positive by construction; configurations in which the coherent $B_RB_\varphi$ correlation reverses sign are outside the adopted closure. We write the vertically integrated radial stress as
\begin{equation}
W=2\alpha\ptot H(1+\delta).
\label{eq:stress_general}
\end{equation}
Here $2\alpha\ptot H$ is the pressure-scaled reference stress and $\delta$
is the fractional stress enhancement associated with net vertical flux. It is
useful to define
\begin{equation}
\alpha_{\rm eff}\equiv\frac{W}{2\ptot H}
=\alpha(1+\delta),
\qquad
\Delta\alpha_{\rm nf}\equiv\alpha\delta.
\label{eq:alphaeff}
\end{equation}

The stress closure is logically separate from the vertical force balance.
Excluding a uniform $p_z$ from the hydrostatic support does not require its
exclusion from the scalar pressure used to normalize the reference MRI
stress. The decomposition in equation~(\ref{eq:stress_general}) is not
unique: in a simulation, part of the variation assigned here to $\delta$
could instead be described as a variation of the effective $\alpha$. The
derivation requires only the equilibrium value of $\delta$ and its
logarithmic thermal response $\etad$. The equivalent criterion written in
terms of the total heating slope is independent of this decomposition.
Section~\ref{sec:results} also repeats the fiducial calculation using
$W_{\rm ref}\propto p_{\rm s}H$.

To translate the net-flux stress enhancement into magnetic-pressure
fractions, we adopt the phenomenological closure
\begin{equation}
\delta=\frac{\zeta\sqrt{\bpol\btor}}{\alpha}.
\label{eq:proxy}
\end{equation}
The coefficient $\zeta$ absorbs the relation between the imposed vertical
field and the coherent radial field that enters the radial Maxwell stress
$-B_RB_\varphi/(4\pi)$. We take $\zeta>0$, corresponding to an enhancement
of outward angular-momentum transport by net vertical flux. Field
configurations producing no additional stress or a sign-changing correlation
are outside this prescription. Equation~(\ref{eq:proxy}) does not identify
$B_zB_\varphi$ directly with the radial Maxwell stress.

For comparison with MRI simulations, we define
\begin{equation}
\zeta_{\rm eff}\equiv
\frac{\Delta\alpha_{\rm nf}}{\sqrt{\bpol\btor}},
\label{eq:zetaeff}
\end{equation}
where $\Delta\alpha_{\rm nf}$ is the increase in the effective stress
parameter relative to a consistently normalized zero-net-flux state.
Equation~(\ref{eq:proxy}) then gives $\zeta_{\rm eff}=\zeta$.

Net-vertical-flux MRI simulations provide qualitative support for the adopted
flux dependence. \citet{BaiStone2013} found that the effective transport
parameter increases from approximately $0.08$ at $\beta_0=10^4$ to values
above unity at $\beta_0=10^2$, where $\beta_0$ is the ratio of midplane gas
pressure to the pressure of the imposed vertical field. \citet{Salvesen2016}
found an approximately inverse-square-root dependence of the effective
transport parameter on $\beta_0^{\rm mid}$ over a broad range of net flux.
At fixed $\btor$, this behavior is consistent with the $\bpol^{1/2}$
dependence in equation~(\ref{eq:proxy}).

These simulations do not provide a direct calibration of $\zeta$. Their
reported stresses contain both Maxwell and Reynolds contributions, their
magnetic fractions are normalized to gas pressure rather than total pressure,
and their thermodynamic regime differs from the radiation-pressure-dominated
disk considered here. They also do not isolate the coherent
$B_RB_\varphi$ contribution or its thermal response. We therefore use
$\zeta=1$ as the fiducial normalization and $\zeta=0.25$ as a weaker-stress
case. These choices are intended to span plausible stress amplitudes rather
than to define empirical bounds.

With $B_z$ and $B_\varphi$ denoting field amplitudes, the additional vertically integrated stress is
\begin{equation}
W_{\rm nf}=2\alpha\ptot H\delta=\frac{\zeta H\left|B_zB_\varphi\right|}{4\pi}.
\label{eq:wnf}
\end{equation}

Because $\zeta$ contains the magnetic geometry and stress efficiency that relate the imposed vertical flux to the radial Maxwell stress, it may also respond during a thermal perturbation. We define
\begin{align}
\etaz&\equiv-\left.\frac{d\ln\zeta}{d\ln H}\right|_{\Sigma,R},\\
\Gnf&\equiv\frac{d\ln W_{\rm nf}}{d\ln H}=1-\gamma-\gp-\etaz.
\label{eq:zeta_response}
\end{align}
Thus $\etad$ measures the response of the fractional correction $\delta$, whereas $\Gnf$ measures the response of $W_{\rm nf}$ itself. For $(\gamma,\gp,\etaz)=(1,0,0)$, $\Gnf=0$, so the additional stress is constant during the idealized expansion while the pressure-scaled reference stress increases. Unless stated otherwise, the quoted magnetic thresholds assume $\etaz=0$.

The local energy equation is
\begin{equation}
\Qvis=\Qrad+\Qadv+\Qwin,
\label{eq:energy}
\end{equation}
with
\begin{align}
\Qvis&=3\alpha\OmegaK\ptot H(1+\delta),
\label{eq:qvis}\\
\Qrad&=\frac{32\sigma T^4}{3\tau},
\qquad
\tau=\frac{\kappa\Sigma}{2},
\label{eq:qrad}\\
\Qadv&=
\frac{\mu_{\rm adv}\Mdot_{\rm acc}\OmegaK^2H^2}{2\pi R^2},
\qquad
\mu_{\rm adv}=1.5.
\label{eq:qadv}
\end{align}
We adopt electron-scattering opacity,
$\kappa=0.34\ {\rm cm^2\,g^{-1}}$.

In equation~(\ref{eq:qrad}), $\Qrad$ is the combined diffusive loss through both disk surfaces in the vertically averaged convention adopted by \citet{Habibi2019}, hereafter H19, and in standard thin-disk treatments \citep{Kato2008}. This normalization does not affect the logarithmic radiative slope $q_{\rm rad}=4$. A factor-of-two normalization test is reported in Section~\ref{sec:results}. The main calculations set $\Qwin=0$.

The wind terms are retained only to display the general extension of the H19 algebra and are not used in the numerical conclusions. The one-zone advective term replaces the radial entropy-gradient dependence by the coefficient $\mu_{\rm adv}$; we therefore vary $1\leq\mu_{\rm adv}\leq2$ when testing the fixed-$B_z$ sequences.

Angular-momentum conservation is represented by
\begin{equation}
\begin{aligned}
\Mdot_{\rm acc}f_{\rm isco}
&=\frac{4\pi\alpha\ptot H(1+\delta)}{\OmegaK},\\
f_{\rm isco}
&=1-\left(\frac{R_{\rm in}}{R}\right)^{1/2}.
\end{aligned}
\label{eq:angmom}
\end{equation}
Equations~(\ref{eq:eos})--(\ref{eq:angmom}) are solved for the local thermal
equilibrium before the perturbation coefficients are evaluated.
\section{Thermal Stability Criterion}
\label{sec:stability}

We perturb the disk at fixed $\Sigma$ and $R$. Define
\begin{align}
N&=4-4\btor-4\bpol-3\bgas,
\label{eq:N}\\
\Cchi&=1-(1-\chi)(1+2\gp)\bpol,
\label{eq:Cchi}\\
\Dchi&=\Cchi+\bgas+2\gamma\btor+2\gp\bpol.
\label{eq:Dchi}
\end{align}
Differentiating the equation of state and the vertical-support relation gives
\begin{equation}
\begin{aligned}
h_T&\equiv
\left.\frac{d\ln H}{d\ln T}\right|_{\Sigma,R}
=\frac{N}{\Dchi},\\
A_T&\equiv
\left.\frac{d\ln\ptot}{d\ln T}\right|_{\Sigma,R}
=\Cchi h_T.
\end{aligned}
\label{eq:responses}
\end{equation}
The derivation is given in Appendix~\ref{app:deriv}.

Using equation~(\ref{eq:Sintro}), the logarithmic heating slope is
\begin{equation}
q_+\equiv\frac{d\ln\Qvis}{d\ln T}
=h_T(\Cchi+1-\Sd).
\label{eq:qplus}
\end{equation}
To retain uncertainty in the advective response, we write $d\ln\Mdot_{\rm acc}=\lmdot\,d\ln W$.

Because the thermal timescale is shorter than the viscous redistribution timescale, the fiducial perturbation holds the local mass flux fixed, $\lmdot=0$. The stress-coupled H19 choice, $\lmdot=1$, is retained as a sensitivity test. Hence $q_{\rm adv}=\lmdot q_++2h_T$, while radiative diffusion gives $q_{\rm rad}=4$. Defining $\fadv=\Qadv/\Qvis$ and the signed wind fraction $\fwin=\Qwin/\Qvis$, the local stability numerator becomes
\begin{align}
\psi=\Dchi\{&(1-\lmdot\fadv)q_+-2\fadv h_T\nonumber\\
&-4(1-\fadv-\fwin)-\fwin q_{\rm w}\}.
\label{eq:psi_expanded}
\end{align}
The mode is unstable for $\psi>0$. On the regular branch,
$\Dchi>0$ and $1-\lmdot\fadv>0$, so the direct thermal-stability condition is
\begin{equation}
\begin{aligned}
&q_+<q_{+,\rm crit},\\
q_{+,\rm crit}
&=\frac{2\fadv h_T+4(1-\fadv-\fwin)+\fwin q_{\rm w}}{1-\lmdot\fadv}.
\end{aligned}
\label{eq:qpluscrit}
\end{equation}
This form is most suitable for comparison with radiation-MHD simulations
because it does not require a decomposition of the turbulent stress. It has
the standard physical interpretation that the logarithmic temperature
response of heating must remain below the corresponding effective cooling
response.

\begin{figure}[t!]
\centering
\includegraphics[width=0.98\columnwidth]{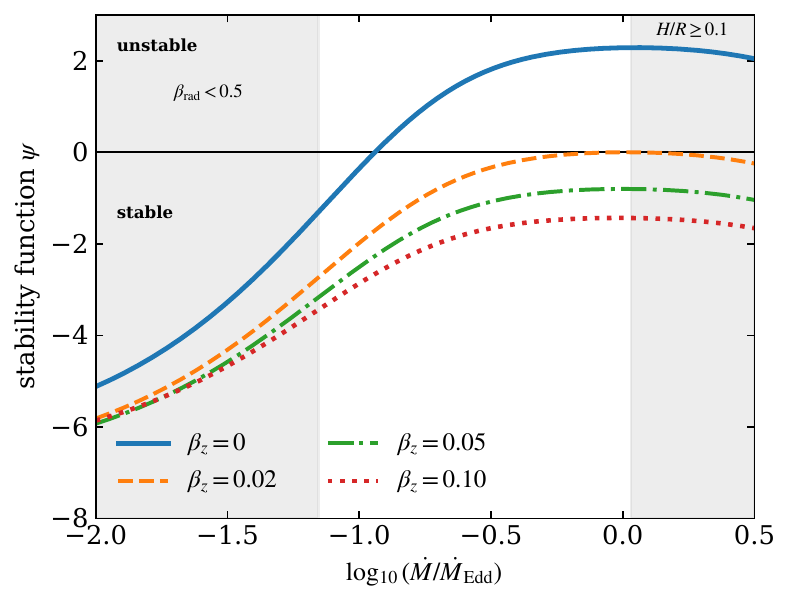}
\caption{Local thermal-stability numerator $\psi$ as a function of accretion rate for $M=10M_\odot$, $R=20\rg$, and $\btor=0.10$ in the wind-free fiducial model. The disk is unstable for $\psi>0$ and stable for $\psi<0$. The shaded bands mark the region of accretion rates for which at least one curve falls outside the adopted radiation-pressure-dominated ($\brad\geq0.5$) or geometrically thin ($H/R<0.1$) domain. The indicated $\bpol$ values are prescribed local equilibrium pressure fractions at each accretion rate; these curves do not conserve $B_z$ and are not fixed-$B_z$ sequences.}
\label{fig:local}
\end{figure}

Using equation~(\ref{eq:qplus}), the criterion is equivalently
\begin{equation}
\Sd>\Sc,
\qquad
\Sc=\Cchi+1-\frac{q_{+,\rm crit}}{h_T}.
\label{eq:Scrit}
\end{equation}
If $\Sc\leq0$, the background disk is already stable and no positive
net-flux response is required. This decomposed form retains the
pressure-scaled reference stress of equation~(\ref{eq:stress_general}), while
the criterion in equation~(\ref{eq:qpluscrit}) remains independent of that
choice.
For the phenomenological stress prescription in equation~(\ref{eq:proxy}),
\begin{equation}
\etad=\Cchi+\gamma+\gp+\etaz,
\label{eq:eta_proxy}
\end{equation}
because $d\ln\ptot=\Cchi\,d\ln H$.

Equivalently, $\etad=(\Cchi+1)-\Gnf$. The stabilizing contribution therefore arises when the additional stress responds more weakly than the pressure-scaled reference stress. Substitution of equations~(\ref{eq:proxy}) and (\ref{eq:eta_proxy}) converts the stress-response condition into a prescription-dependent $\bpol^{\rm crit}$. At $\bpol=0$, $\fwin=0$, and $\lmdot=1$, equation~(\ref{eq:psi_expanded}) recovers the wind-free toroidal-field result of H19. Removing all magnetic, advective, and wind terms gives $\psi=4-10\bgas$.

\section{Results}
\label{sec:results}

\begin{figure*}[!tp]
\centering
\includegraphics[width=0.8\textwidth]{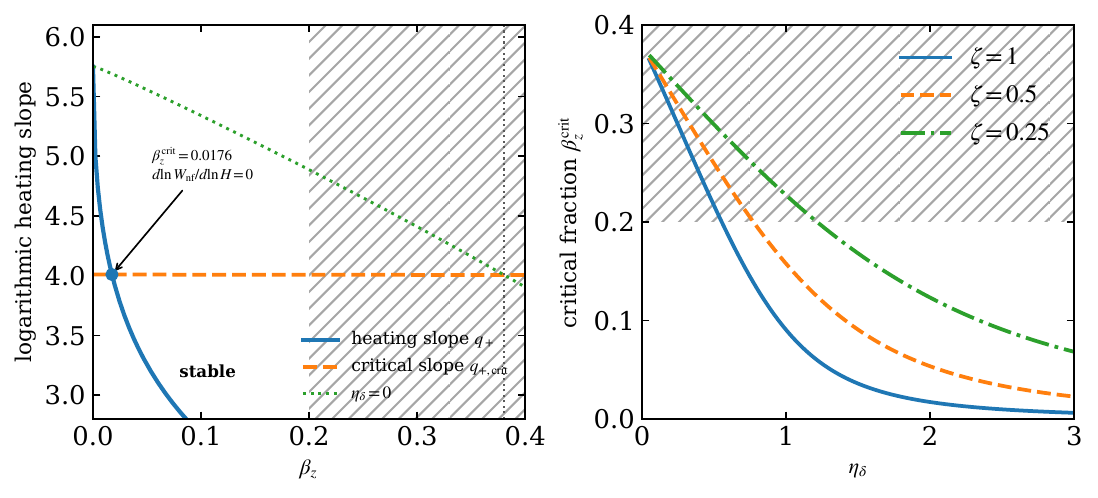}
\caption{Direct local thermal-stability criterion for $M=10M_\odot$, $\Mdot=0.5\MEdd$, and $\btor=0.10$. Left: the disk is stable where $q_+<q_{+,\rm crit}$, and the fiducial closure gives $\bpol^{\rm crit}=0.0176$ with $\Gnf=0$. The dotted curve is the diagnostic $\etad=-d\ln\delta/d\ln H=0$; it holds the fractional correction $\delta$ fixed and therefore makes $W_{\rm nf}$ track the logarithmic response of the reference stress. Its formal crossing at $\bpol\simeq0.381$ is not a zero-response limit for $W_{\rm nf}$ and lies outside $\beta_{\rm mag}<0.3$. Right: critical vertical magnetic-pressure fraction as a function of the fractional-correction response exponent $\etad$ for three values of $\zeta$.}
\label{fig:response}
\end{figure*}

\begin{figure*}[!tp]
\centering
\includegraphics[width=0.98\textwidth]{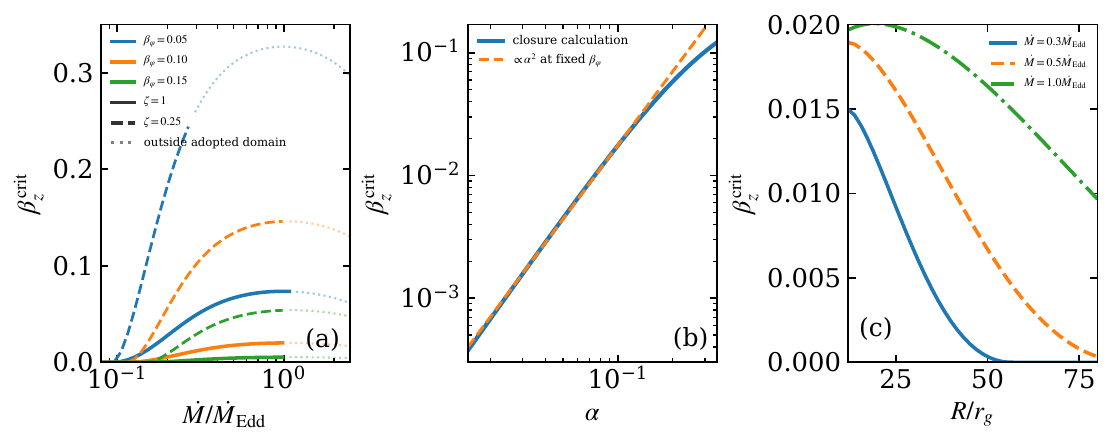}
\caption{Critical vertical magnetic-pressure fraction $\bpol^{\rm crit}$ inferred from the adopted net-flux stress closure for a $10M_\odot$ black-hole accretion disk. Panel (a) shows the dependence on accretion rate for prescribed local toroidal magnetic-pressure fractions $\btor=0.05$, $0.10$, and $0.15$, with $\zeta=1$ and $0.25$. Faint dotted portions lie outside $\brad\geq0.5$, $\tau\geq10$, $H/R<0.1$, or $\beta_{\rm mag}<0.3$. These curves vary $\btor$ at fixed values of the remaining closure parameters and therefore represent sensitivity tests rather than self-consistent MRI saturation sequences. Panel (b) shows the dependence on $\alpha$ at $\Mdot=0.5\MEdd$ and $\btor=0.10$. The dashed $\bpol^{\rm crit}\propto\alpha^2$ curve is the algebraic scaling implied by the adopted closure, not a universal MRI prediction. Panel (c) shows the radial variation for $\Mdot=0.3$, $0.5$, and $1.0\,\MEdd$ under the fiducial closure.}
\label{fig:boundary}
\end{figure*}

Unless otherwise stated, the local calculation uses $M=10M_\odot$, $R=20\rg$, $\alpha=0.1$, $\btor=0.10$, $\gamma=1$, $\gp=0$, $\etaz=0$, $\chi=0$, $\zeta=1$, $\lmdot=0$, $\mu_{\rm adv}=1.5$, and $\fwin=0$. Parameter variations in Figs.~\ref{fig:local}, \ref{fig:response}, \ref{fig:boundary}, and \ref{fig:closure} are one-at-a-time closure sensitivity tests rather than independent MRI saturation sequences.

Figure~\ref{fig:local} establishes how the local thermal mode changes across accretion rate when the magnetic-pressure fractions are prescribed independently at each equilibrium point. To isolate the physical origin of this stabilization more directly, we now fix the disk at the representative state $M=10M_\odot$, $R=20\rg$, and $\Mdot=0.5\MEdd$ and examine how the temperature dependence of the heating changes as the net-vertical-flux contribution to the stress is varied. This comparison separates the direct thermal-stability condition from the closure-dependent conversion of that condition into a critical vertical magnetic-pressure fraction.

\begin{table*}[!tp]
\centering
\caption{iducial marginal states for $M=10M_\odot$, $\Mdot=0.5\MEdd$, $R=20\rg$, $\btor=0.10$, $\lmdot=0$, and $\etaz=0$.}
\label{tab:marginal}
\begin{tabular}{cccccccccc}
\hline\hline
$\zeta$ & $\bpol^{\rm crit}$ & $\bgas$ & $\beta_{\rm mag}$ & $p_z/p_{\rm gas}$ & $\delta$ & $\alpha_{\rm eff}$ & $H/R$ & $\tau$ & $\fadv$\\
\hline
1.00 & 0.0176 & 0.0231 & 0.1176 & 0.761 & 0.419 & 0.142 & 0.0482 & 1505 & 0.00514\\
0.25 & 0.1338 & 0.0195 & 0.2338 & 6.86 & 0.289 & 0.129 & 0.0489 & 1419 & 0.00529\\
\hline
\end{tabular}
\end{table*}

Figure~\ref{fig:response} presents this comparison. In the left panel, the logarithmic heating slope $q_+$ is shown together with the critical slope $q_{+,\rm crit}$. For the fiducial closure, $(\gamma,\gp,\etaz)=(1,0,0)$, the additional net-flux stress satisfies $\Gnf=d\ln W_{\rm nf}/d\ln H=0$, and marginal stability occurs at $\bpol^{\rm crit}=0.0176$. The dotted curve is a diagnostic construction in which the fractional correction is held fixed during the perturbation, $\etad=-d\ln\delta/d\ln H=0$. In this case $W_{\rm nf}$ follows the same logarithmic response as the pressure-scaled reference stress, and the formal crossing shifts to $\bpol\simeq0.381$, outside the adopted $\beta_{\rm mag}<0.3$ domain.

The right panel of Fig.~\ref{fig:response} isolates the dependence on the fractional-correction response itself. Treating $\etad$ as an independent parameter, the inferred $\bpol^{\rm crit}$ is shown for several values of the stress-amplitude parameter $\zeta$. A larger $\etad$ means that the fractional net-flux contribution decreases more rapidly as the disk expands, which weakens the temperature dependence of the total heating and lowers the vertical magnetic fraction required for marginal stability. The figure therefore demonstrates that $\bpol^{\rm crit}$ is determined jointly by the equilibrium stress enhancement and its thermal response, rather than by the magnetic-pressure fraction alone.

Table~\ref{tab:marginal} reports the magnetic fractions in both total-pressure and gas-pressure normalization. The $\zeta=1$ marginal state has $p_z/p_{\rm gas}=0.76$, while the $\zeta=0.25$ state has $p_z/p_{\rm gas}=6.86$. The latter is therefore strongly magnetized relative to the gas pressure even though it satisfies the adopted total-pressure cut $\beta_{\rm mag}<0.3$.

Figure~\ref{fig:boundary} gives the closure-dependent mapping from the local response criterion to $\bpol^{\rm crit}$. At the fiducial point,
\begin{equation}
\bpol^{\rm crit}=0.0176\quad(\zeta=1),
\qquad
\bpol^{\rm crit}=0.1338\quad(\zeta=0.25),
\label{eq:zeta_numbers}
\end{equation}
for $\etaz=0$. Panel (a) varies the prescribed local $\btor$ while the remaining closure parameters are held fixed. Because imposed vertical flux, toroidal magnetic energy, and turbulent stress are correlated in MRI turbulence, these curves are sensitivity tests rather than self-consistent MRI sequences. Panel (b) similarly isolates the algebraic $\alpha$ dependence. At approximately fixed marginal stress enhancement,
\begin{equation}
\bpol^{\rm crit}\simeq\frac{\alpha^2\delta_{\rm crit}^2}{\zeta^2\btor},
\label{eq:alpha_scaling}
\end{equation}
so the near-$\alpha^2$ behavior follows from the adopted closure and is not a universal prediction of MRI turbulence. Panel (c) shows the radial threshold. At lower accretion rates $\bpol^{\rm crit}$ decreases outward as the gas-pressure contribution grows, whereas the near-Eddington sequence develops a broad maximum where radiation pressure remains strong but advective stabilization has weakened.

\begin{table*}[t]
\centering
\caption{Sensitivity tests at the fiducial radius and accretion rate. We fix, $\lmdot=0$ and $\etaz=0$, unless listed otherwise.}
\label{tab:sensitivity}
\begin{tabular}{lcc}
\hline\hline
Parameter & Values & $\bpol^{\rm crit}$\\
\hline
$\alpha$ & $0.02,\ 0.10,\ 0.30$ &  $6.8\times10^{-4},\ 0.0176,\ 0.102$\\
$\gamma$ & $0.50,\ 0.75,\ 1.00,\ 1.25,\ 1.50$ & $0.0660,\ 0.0327,\ 0.0176,\ 0.0100,\ 0.00587$\\
$\gp$ & $0.00,\ 0.25,\ 0.50,\ 0.75,\ 1.00$ & $0.0176,\ 0.0127,\ 0.00957,\ 0.00749,\ 0.00603$\\
$\zeta$ & $0.25,\ 0.50,\ 1.00$ & $0.1338,\ 0.0566,\ 0.0176$\\
$\etaz$ & $-0.50,\ 0.00,\ 0.50$ & $0.0383,\ 0.0176,\ 0.00987$\\
$\lmdot$ & $0.00,\ 1.00$ & $0.0176,\ 0.0170$\\
$\mu_{\rm adv}$ & $1.0,\ 1.5,\ 2.0$ & $0.0177,\ 0.0176,\ 0.0175$\\
Radiative coefficient & $1.00,\ 0.50$ & $0.0176,\ 0.0202$\\
Reference pressure & $\ptot,\ p_{\rm s}$ & $0.0176,\ 0.0180$\\
\hline
\end{tabular}
\end{table*}

\begin{figure*}[!tp]
\centering
\includegraphics[width=0.90\textwidth]{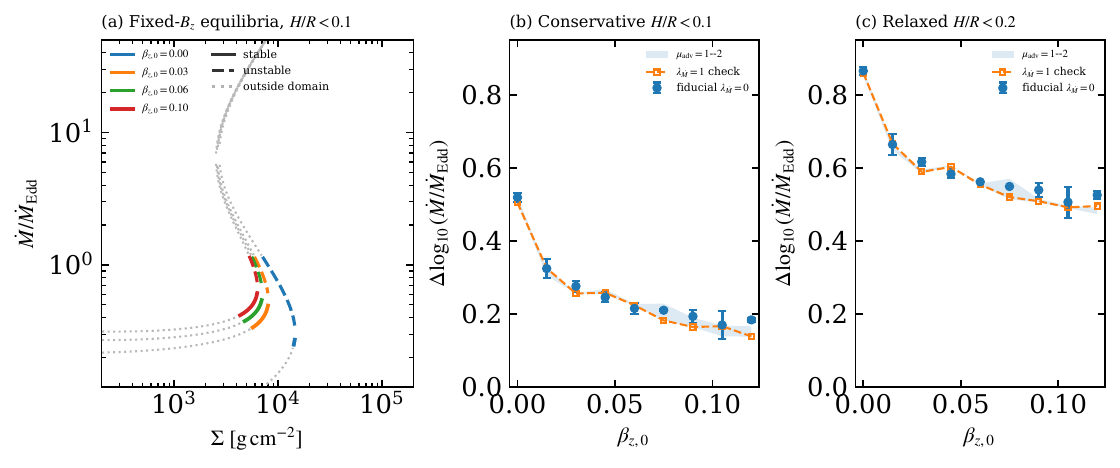}
\caption{Fixed-$B_z$ equilibrium sequences and the thermally unstable accretion-rate interval for a $10M_\odot$ disk. The equilibrium family uses $B_\varphi H^{\gammaeq}={\rm const}$ with $\gammaeq=1$, whereas the local thermal derivative uses $\gamma=1$. Panel (a) shows stable and unstable segments for $\beta_{z,0}=0$, $0.03$, $0.06$, and $0.10$ under $H/R<0.1$; grey dotted portions lie outside the adopted domain. Panels (b) and (c) show $\Delta\log_{10}(\Mdot/\MEdd)$ for $H/R<0.1$ and $H/R<0.2$. Filled points use $\lmdot=0$, the shaded band spans $1\leq\mu_{\rm adv}\leq2$, and open squares show the $\lmdot=1$ test. The unstable width decreases from $0.52$ to $0.18$ dex for $H/R<0.1$ and from $0.87$ to $0.53$ dex for $H/R<0.2$. Vertical brackets indicate the two-grid numerical difference, not statistical errors.}
\label{fig:scurve}
\end{figure*}

\begin{figure*}[!tp]
\centering
\includegraphics[width=0.90\textwidth]{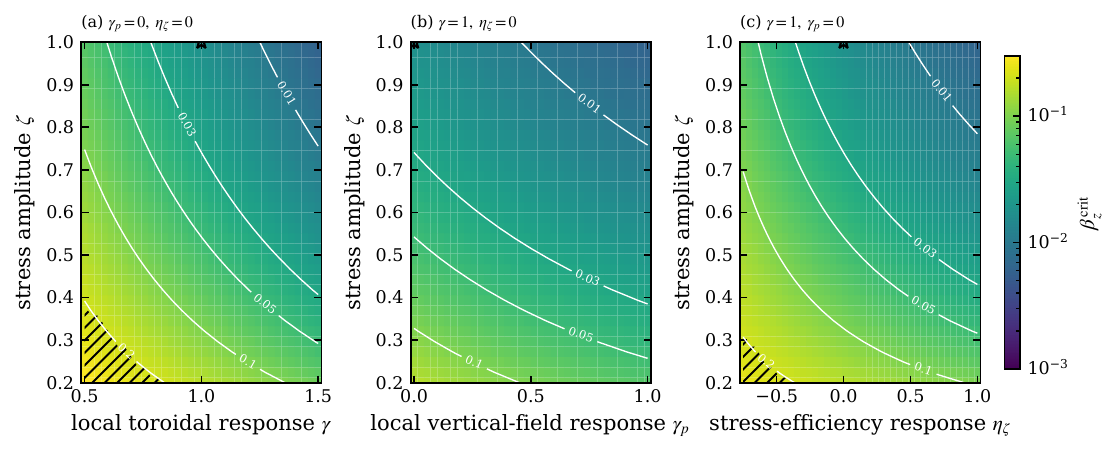}
\caption{Closure dependence of $\bpol^{\rm crit}$ for $M=10M_\odot$, $\Mdot=0.5\MEdd$, $R=20\rg$, and $\btor=0.10$. Panel (a) varies the local toroidal response $\gamma$ and $\zeta$ for $\gp=0$ and $\etaz=0$; panel (b) varies $\gp$ and $\zeta$ for $\gamma=1$ and $\etaz=0$. Panel (c) varies the stress-efficiency response $\etaz=-d\ln\zeta/d\ln H$ and $\zeta$ for $\gamma=1$, $\gp=0$. Positive $\etaz$ weakens the response of the additional stress during expansion and lowers the required $\bpol^{\rm crit}$. White contours show selected thresholds, stars mark the fiducial closure, and hatched regions satisfy $\bpol^{\rm crit}+\btor\geq0.3$.}
\label{fig:closure}
\end{figure*}

A local thermal derivative does not determine how the saturated toroidal field varies between distinct steady equilibria. We therefore construct the fixed-$B_z$ families with the separate equilibrium closure $B_\varphi H^{\gammaeq}={\rm const}$ and adopt $\gammaeq=1$ for this illustrative calculation. Local stability at every equilibrium point is evaluated independently with $\gamma=1$. Figure~\ref{fig:scurve}(a) applies the same conservative $H/R<0.1$ cut used in the local analysis.

Under this strict thickness criterion, the unstable width decreases from approximately $0.52$ dex at $\beta_{z,0}=0$ to $0.18$ dex at $\beta_{z,0}=0.12$ (see Fig.~\ref{fig:scurve}(b)). The difference between the two numerical grids is at most $0.04$ dex. Repeating the calculation with $H/R<0.2$ gives a decrease from approximately $0.87$ to $0.53$ dex [Fig.~\ref{fig:scurve}(c)]. Reclassifying the same equilibria with $\lmdot=1$ instead of the fiducial $\lmdot=0$ leaves the sampled widths unchanged at the quoted resolution, while $1\leq\mu_{\rm adv}\leq2$ produces only a narrow envelope in the conservative domain. The contraction of the unstable interval is therefore present under both thickness cuts, although its quantitative width remains conditional on the equilibrium and one-zone advective prescriptions.

Figure~\ref{fig:closure} isolates the remaining uncertainty in the local magnetic response. Panels (a) and (b) vary $\gamma$ and $\gp$ at fixed $\etaz=0$. Panel (c) introduces $\etaz=-d\ln\zeta/d\ln H$, which describes thermal changes in the magnetic geometry and stress efficiency absorbed into $\zeta$. Positive $\etaz$ makes $\zeta$ decrease during expansion and weakens the response of $W_{\rm nf}$, whereas negative $\etaz$ has the opposite effect. At the fiducial point, varying $\etaz$ from $-0.5$ to $0.5$ moves $\bpol^{\rm crit}$ from $0.0383$ to $0.00987$.

\FloatBarrier

\section{Discussion}
\label{sec:discussion}

The analysis separates the equilibrium magnetization from the thermal response of the stress associated with net vertical flux. The total heating slope $q_+$ gives a local criterion that is independent of how the turbulent stress is decomposed, while $\Sd$ is useful when the stress is written as pressure-scaled and net-flux components. In this decomposition $\etad$ characterizes the response of the fractional correction $\delta$, whereas $\Gnf$ characterizes the response of $W_{\rm nf}$. For the fiducial closure $\Gnf=0$; by contrast, setting $\etad=0$ makes the additional stress track the reference stress.

Earlier analytical treatments showed that an MRI-generated toroidal field can reduce the temperature sensitivity of the heating rate \citep{Zheng2011,Habibi2019}. The present model adds ordered vertical flux through its contribution to the radial stress rather than treating $B_z$ only as an additional pressure term. The conversion from the robust heating-slope criterion to $\bpol^{\rm crit}$ remains closure dependent because it requires $\zeta$, $\gamma$, $\gp$, and $\etaz$. The one-at-a-time scans in Figs.~\ref{fig:boundary} and \ref{fig:closure} should therefore be read as sensitivity calculations, not as dynamically self-consistent MRI sequences.

Strong magnetic fields can alter the vertical structure of the disk and weaken the classical radiation-pressure instability \citep{Machida2006,Oda2009,BegelmanPringle2007,Sadowski2016,DexterBegelman2019}. In the present calculation, local stabilization does not require the vertical magnetic pressure to dominate the hydrostatic support. Table~\ref{tab:marginal} nevertheless shows that a small $\bpol$ in total-pressure normalization need not correspond to weak magnetization relative to the gas pressure. The $\zeta=0.25$ marginal state has $p_z/p_{\rm gas}=6.86$, so comparisons with net-flux MRI calculations must use a consistent pressure normalization.

The same distinction is relevant to radiation-MHD simulations. \citet{Hirose2009} obtained long-lived radiation-dominated states in which stress fluctuations preceded pressure fluctuations, whereas \citet{Jiang2013} found thermal expansion or collapse in larger shearing boxes. The condition $q_+<q_{+,\rm crit}$ is therefore most directly applicable to a controlled low-frequency thermal response. A turbulent time-series correlation can contain phase lags, frequency dependence, and nonlocal transport and is not automatically the derivative used in the local stability calculation.

The fixed-$B_z$ sequences address the topology of an equilibrium family and require an additional assumption. The relation $B_\varphi H^{\gammaeq}={\rm const}$ describes how the saturated toroidal field is assigned between distinct steady states, whereas $\gamma$ describes a local thermal derivative. The contraction of the unstable interval in Fig.~\ref{fig:scurve} is consequently a result for the adopted $\gammaeq=1$ family. Its persistence for $H/R<0.1$ and $H/R<0.2$, for $\lmdot=0$ and $1$, and over $1\leq\mu_{\rm adv}\leq2$ shows that the qualitative trend is not produced solely by the previous thickness or advective choices, but it does not make the trend universal.

Applications to BHXB limit cycles and changing-look AGN remain speculative. The model does not evolve surface density and magnetic flux together and does not predict transition times, nonlinear amplitudes, recurrence times, or duty cycles. In AGN, additional opacity physics can also modify vertical transport and thermal stability \citep{Jiang2016Iron}.

The calculation is local, linear, vertically integrated, Newtonian, and restricted to electron-scattering opacity. It determines the sign of the thermal response but not the growth rate and does not address the secular Lightman--Eardley viscous mode \citep{Lightman1974}. The one-zone advective term replaces the radial entropy gradient by $\mu_{\rm adv}$, the magnetic response is prescribed rather than derived from an induction equation, and the positive $W_{\rm nf}$ closure excludes sign reversals of the coherent Maxwell correlation. The wind terms are retained only in the general algebra and are not used as numerical evidence for the main conclusions.

\section{Conclusions}
\label{sec:conclusions}

We have derived a local thermal-stability criterion for a radiation-pressure-dominated accretion disk in which net vertical magnetic flux contributes to the radial stress. The direct condition is $q_+<q_{+,\rm crit}$. In the pressure-scaled stress decomposition, $\Sd=[\delta/(1+\delta)]\etad$ measures the response of the fractional stress correction, while $\Gnf=d\ln W_{\rm nf}/d\ln H$ measures the response of the additional stress itself. For the illustrative closure $\delta=\zeta\sqrt{\bpol\btor}/\alpha$, $\etad=\Cchi+\gamma+\gp+\etaz$ and $\Gnf=1-\gamma-\gp-\etaz$. With $\lmdot=0$ and $(\gamma,\gp,\etaz)=(1,0,0)$, the fiducial $10M_\odot$ disk gives $\bpol^{\rm crit}=0.0176$ for $\zeta=1$ and $0.1338$ for $\zeta=0.25$. These are conditional closure values rather than universal magnetic thresholds.

Fixed-$B_z$ equilibrium sequences were recalculated with a separate equilibrium index $\gammaeq$, the conservative $H/R<0.1$ cut, the relaxed $H/R<0.2$ cut, $\lmdot=0$ and $1$, and $1\leq\mu_{\rm adv}\leq2$. For the adopted $\gammaeq=1$ family, increasing the reference $B_z$ narrows but does not remove the unstable interval. The quantitative width remains conditional on the assumed equilibrium and advective prescriptions.

The simulation-facing result is therefore the total heating response. Measuring the low-frequency $q_+$, together with the cooling and advective responses, provides a direct test of the local criterion; measurements of $\delta$, $\zeta$, $\etad$, and $\etaz$ can then determine whether the specific net-flux closure accounts for that response.

\section*{Acknowledgments}

We thank the anonymous reviewer for improving the draft substantially. The authors thank Maryam Samadi for valuable scientific comments and for a careful reading of the manuscript.

\software{NumPy \citep{Harris2020}, SciPy \citep{Virtanen2020},
Matplotlib \citep{Hunter2007}}

\section*{DATA AND CODE AVAILABILITY}

The numerical calculations and figure-generation material used in this work are publicly available through the project GitHub repository at \url{https://github.com/magnetosamik/Thermal-stability-net-vertical-flux}. The manuscript-specific release is archived on Zenodo \citep{Mitra2026zenodo} at \dataset[doi:10.5281/zenodo.22052047]{https://doi.org/10.5281/zenodo.22052047}. The archived material contains the source and executed Jupyter notebook used to generate Figs.~1--5, together with the machine-readable fiducial marginal states, the sensitivity calculations reported in Table~2, and the fixed-$B_z$ unstable-width calculations for both $H/R<0.1$ and $H/R<0.2$, including the $\lmdot$ and $\mu_{\rm adv}$ tests. The accompanying README maps the numerical material to the corresponding figures, tables, and quoted numerical results and provides the required software dependencies and execution instructions. No observational or proprietary data are used in this work.

\bibliographystyle{aasjournal}
\bibliography{references}

\appendix

\section{General Perturbation Derivation}
\label{app:deriv}

At fixed $\Sigma$, let
$h=d\ln H/d\ln T$ and $A=d\ln\ptot/d\ln T$. Differentiating equation~
(\ref{eq:support}) gives
\begin{equation}
\left[1-(1-\chi)\bpol\right]h
=A+2(1-\chi)\gp\bpol h,
\end{equation}
and hence
\begin{equation}
A=\Cchi h,
\qquad
\Cchi=1-(1-\chi)(1+2\gp)\bpol.
\label{eq:appC}
\end{equation}
The equation of state gives
\begin{equation}
A=N-(\bgas+2\gamma\btor+2\gp\bpol)h.
\end{equation}
Combining these expressions yields equation~(\ref{eq:responses}).

For a general stress correction,
\begin{equation}
d\ln(1+\delta)
=-\frac{\delta}{1+\delta}\etad\,d\ln H
=-\Sd h\,d\ln T.
\end{equation}
Since $\Qvis\propto\ptot H(1+\delta)$, this gives equation~
(\ref{eq:qplus}). We parameterize the local mass-flux response as
$d\ln\Mdot_{\rm acc}=\lmdot d\ln W$, so
$\Qadv\propto\Mdot_{\rm acc}H^2$ gives
$q_{\rm adv}=\lmdot q_++2h_T$. Substituting these slopes into the perturbed
energy balance gives equation~(\ref{eq:psi_expanded}). Solving $\psi=0$ for
$\Sd$ gives equation~(\ref{eq:Scrit}).

For the phenomenological stress prescription,
\begin{align}
d\ln\delta&=d\ln\zeta+d\ln B_z+d\ln B_\varphi-d\ln\ptot\\
&=-(\Cchi+\gamma+\gp+\etaz)d\ln H,
\end{align}
which proves equation~(\ref{eq:eta_proxy}). Separately, $d\ln W_{\rm nf}/d\ln H=1-\gamma-\gp-\etaz=\Gnf$; thus $\etad$ and $\Gnf$ are different derivatives. Setting $\bpol=0$, $\fwin=0$, and $\lmdot=1$ eliminates $\delta$ and recovers the wind-free \citetalias{Habibi2019} result to machine precision in the numerical implementation. Removing all magnetic, advective, and wind terms gives $\psi=4-10\bgas$.

\section{Optional Parametric Wind}
\label{app:wind}

The general equations retain a signed wind fraction $\fwin$. The following prescription is retained only for algebraic continuity with H19. We use $s=1/2$, $R_{\rm d}=200\rg$, and denote the supplied outer rate by $\Mdot_{\rm out}$:
\begin{align}
\Mdot_{\rm acc}(R)&=\Mdot_{\rm out}\left[1-s_{\rm w}
\left(1-\frac{R^s-R_{\rm in}^s}{R_{\rm d}^s-R_{\rm in}^s}\right)\right],\\
\dot m_{\rm w}(R)&=\frac{s s_{\rm w}\Mdot_{\rm out} R^{s-2}}{4\pi(R_{\rm d}^s-R_{\rm in}^s)}.
\end{align}
The wind energy term is
\begin{equation}
\Qwin=\frac{1}{2}\dot m_{\rm w}
(\eta_{\rm b}f_v^2+\eta_{\rm k})\OmegaK^2R^2,
\qquad f_v=\sqrt{2},
\end{equation}
with the coefficients of \citetalias{Habibi2019}:
$\eta_{\rm b}=3-2l^2$ and $\eta_{\rm k}=1$ for $l^2<3/2$, while
$\eta_{\rm b}=0$ and $\eta_{\rm k}=5/2-l^2$ for $l^2>3/2$.
During the local thermal perturbation we hold the supplied outer rate
$\Mdot_{\rm out}$ fixed. Equations above then give
$q_{\rm w}=d\ln\Qwin/d\ln T=0$ at fixed $R$. A wind whose mass loading follows
the local stress would require an additional response law and is not included.
All numerical results in this paper set $s_{\rm w}=0$; the wind extension is not used as evidence for the main conclusions.

\section{Numerical Implementation Checks for the Fixed-$B_z$ Curves}
\label{app:verification}

For the fixed-$B_z$ sequences, the equilibrium residual is
$\mathcal{L}=\Qvis-\Qrad-\Qadv$ with $\Qwin=0$. At every root we compare the
analytical derivative $\psi/\Dchi$ with
\begin{equation}
\frac{1}{\Qvis}\frac{d\mathcal{L}}{d\ln T}
\simeq
\frac{\mathcal{L}(Te^{\epsilon})-\mathcal{L}(Te^{-\epsilon})}{2\epsilon\Qvis},
\qquad \epsilon=10^{-5}.
\end{equation}
This finite-difference comparison is performed only for the implementation-check case $\gammaeq=\gamma=1$, $\lmdot=1$, and $\etaz=0$, for which the numerical perturbation follows the same response law as the analytical derivative. The maximum absolute discrepancy is $5.92\times10^{-8}$ and the stability signs agree at every point. This verifies the coded derivative but does not independently validate the physical closure or identify $\gammaeq$ with $\gamma$. The same
notebook verifies
$|\psi(\bpol=0)-\psi_{\rm H19}|=2.22\times10^{-16}$ and recovers the classical
radiation-pressure value $\psi=4$ when $\bgas=0$ and all magnetic, advective,
and wind terms are removed. The thin-disk interval in Fig.~\ref{fig:local}
has $0.073\leq\Mdot/\MEdd\leq1.034$ and minimum $\tau=499$ across all four
curves. The valid displayed fixed-$B_z$ segments have minimum $\tau=574$ and
minimum $\brad=0.682$.

\end{document}